\documentclass[preprint,onecolumn, aps, floats,showpacs, nofootinbib]{revtex4}
\usepackage{amsmath} 
\usepackage{graphicx}
\begin{document}
\title{\bf A covering probability amplitude formalism for classical and quantum mechanics, and possible new modes of non-classical behaviour}


\author{ Ram K. Varma}
\email{ramkvarma@gmail.com\\
fax: +91-79-26314460\\
phone:+91-79-26749242}

\affiliation{ Physical Research Laboratory, Navrangpura, Ahmedabad 380 009, \ \
INDIA}
\begin{abstract}
\noindent

A generalized dynamics is postulated in a product space ${\cal R}^{3}\times {\cal S}^{1}$ with ${\cal R}^3$ representing the configuration space of a one particle system to which is attached the $U(1)$ fibre bundle represented by the manifold ${\cal S}^{1}$.
The manifold ${\cal S}^{1}$ is chosen to correspond to the action phase, with the action  being the principal function corresponding to the associated classical dynamical system. A Hilbert space representation of the flow equation representing the dynamics in the above product space is then obtained, which is found to yield a probability amplitude formalism in terms of a generalized set of equations of the Schr\"odinger form which constitutes a covering formalism for both quantum and classical mechanics. Being labelled by an integral index $n$, the equation corresponding to $n=1$ is identified with the Schr\"odinger equation, while equations corresponding to $n\rightarrow large$ yield the classical dynamics through the WKB limit. Equations with lower indices $ n=2, \ 3, \ 4,...$ predict the existence of new modes of non-classical behaviour, the observational implications of which are discussed. Some preliminary experimental results are presented which point to the existence of these modes of non-classical behaviour.

\end{abstract}

\maketitle

{\bf 1. Introduction}

 There have been numerous attempts--too numerous to quote them all-- since the advent of quantum mechanism to fathom the precise
 relationship between the latter and its `classical sister' --the classical mechanics. Though much understanding has been achieved, there are yet many aspects which remain unresolved.
In recent years, a program involving the concept of decoherence has been developed which seeks to connect the state vector description of quantum mechanics with the classicality of the outcomes of measurement. The `measurement problem' of the earlier years is thus sought to be given a new outlook through the decoherence program. Proposed initially by Zurek$^{1}$, wherein a very informative overview of its basic concepts has been presented by him, while a very comprehensive recent review of the subject has been presented by Schlosshauer$^{2}$

While these are undoutedly important questions which need attention for a better understanding of the persisting foundational issues in the framework of the current theory, we address here a somewhat different question. The question relates to the very nature of probability in quantum mechanics vis-a-vis classical probability, and which impinches on the issue of the so-called `incompleteness' of the quantum description. The apparently rather paradoxical aspect of quantum mechanics as posited against classical mechanics, as is well known, is that while the former is a more general theory which includes the latter as a limit, it is nevertheless intrinsically probabilistic as against the deterministic 
nature of the latter.  Moyal ~\cite{moyal} has, in fact, developed what he refers to as `Quantum Mechanics as a Statistical 
 Theory', whereby he tries to bring out its inherent probabilistic character in explicit terms. He has obtained an evolution equation for a joint probability (phase space) distribution for the 
 eigenvalues of non-commuting observables $p$ and $q$. Moyal's formalism thus attempts to cast quantum mechanics in terms of functions which are like Liouville density functions obeying Liouville-like equations so that the probabilities in quantum mechanics could be viewed in terms of classical looking probabilities in terms of the phase space functions. It was though Wigner~\cite{wigner} who had first introduced such distribution functions of `position' and `momentum', referred to Wigner functions.
But, as is well known, such distributions do not represent genuine probabilities because of their non-positive nature over the entire
region of phase space so defined. Thus despite its appearance as a  Liouville-like equation, the Moyal evolution equation does not represent a phase flow.  Furthermore, it still represents a boundary value problem since, not unexpectedly, it is equivalent to the Schr\"odinger equation as shown in Appendix 4 of Moyal's paper ~\cite{moyal}

Even though the Wigner-Moyal distributions use `momentum' as a variable, identifying it with the Fourier transform variable $ k $, it is not a genuine momentum independent of the coordinate $x$ in the same sense that it appears in the classical phase space. This `psuedo-momentum' $k$ represents essentially the inverse of the same configuration space over which the Schr\"odinger wave function is defined; it corresponds to the finer scale variation of the wave function, when the scales are well separated, but ill-defined as a momentum when there is no separation of scales. It is this fact which would appear to be responsible for the non-positive character of these distributions, and related eventually to the uncertainty principle. However, it is this non-positivity which represents the essence of wave amplitude description of quantum mechanics and the interference effects characteristic of the quantum behaviour.

In view of the above observations, we explore a different point of view with respect to the nature of the non-positivity of these distributions, which essentially have
Schr\"odinger wave amplitudes as their basis. We pose the question whether the intrinsic probabilistic nature of the Schr\"odinger description and the non-positivity of the Wigner-Moyal distributions is ultimately related to, what has been regarded since Einstein, the `incompleteness' of the quantum description. We thus seek to supplement the Schr\"odinger amplitude, calling it $\Psi^{(1)}$, by other amplitudes $\Psi^{(n)}, n=2,\ 3, \ 4,...$, so that together they represent a complete set belonging to a $U(1)$ fibre bundle, 
and transform according to the irreducible representation of $U(1)$. The fibre bundle $U(1) $ is then attached to the space $\mathcal{R}^{3}\times \mathcal{C}$; the latter manifold being the natural home of the Schr\"odinger wave functions.
The total space is then 
$ E = \mathcal{R}^{3}\times \mathcal{C}\times\mathcal{S}^{1}$, where $\mathcal{S}^{1}$ represents the circle. As a manifold, $U(1)$ is just the circle $\mathcal{S}^{1}$.

Accordingly, we define a deterministic dynamics in the higher dimensional space $\mathcal{R}^3\times \mathcal{S}^{1}$ which is a generalization of the classical Hamiltonian dynamics, but one which will be seen to be intimately related to the Hamilton-Jacobi version thereof. A generalized Liouville type equation can then be written down for this generaized dynamics. This will be done in the next Section 2.  In Section 3, we shall carry out what may be regarded as a Hilbert space representation of this generalized Liouville equation, and will show that it leads to a generalized (infinite) set of Schr\"odinger equations, whose wave functions belong to the vector space defined on the unit circle $\mathcal{S}^{1}$. These set of Schr\"odinger equations are thus labelled by the integers $n$ so that now these Schr\"odinger wave amplitudes $\Psi^{(n)}$ belong to the irreducible representation of the $U(1)$ group, and thus form a complete set of which Schr\"odinger wave function is just one of the elements corresponding to $n=1$.

We first develop this point of view essentially mathematically, and will later relate it to the actual physical issue of quantum mechanics, which will be seen to be contained in this formalism as just one of the modes-$n$. We shall discuss, how quantum mechanics and known quantum behaviour of systems can be related to this formalism, and how the former may be looked at from this perspective. For instance, when applied to the problem of the double-slit interference, it is shown in the subsequent section, that when the probability density at the observation screen is summed over all the modes $n$ with equal weights, the interference term disappears. This leads to the conclusion that the selection of only $n=1$ mode out of the infinite ones, is responsible for the existence of interference effects, and the Schr\"odinger formalism which describes the latter is, in this picture, thus incomplete in this sense. Its probabalistic character may, in this perspective, be traced to this incompleteness.

 {\bf 2. A generalized Liouville equation in an extended space $\mathcal{R}^{3}\times \mathcal{S}^{1}$ }

We first begin by writing the classical Liouville equation, which is given as

\begin{equation}
\frac{\partial f}{\partial t} + {\dot {\bf x}}\cdot\frac{\partial f}{\partial {\bf x}}+ {\dot{\bf p}}\cdot
\frac{\partial f}{\partial {\bf p}} = 0,
\label{eq.(1)}
\end{equation}
which evolves along its characteristics given by

\begin{equation}
{\dot{\bf p}}=  -\frac{\partial H}{\partial {\bf x}} ; \   \   \  {\dot {\bf x}} = \frac{\partial H}
{\partial {\bf p}},
\label{eq.(2)}
\end{equation}
with the Hamiltonian being

\begin{equation}
H = \frac{p^2}{2m} + V({\bf x}, t),
\label{eq.(3)}
\end{equation}
 or any other more general form.

A solution of the Liouville equation can be written down in terms of the momentum and position constants of motion $(\alpha, \beta)$, which correspond to the vanishing of the transformed `new' Hamiltonian in a canonical transformation: $f= f(\alpha, \beta)$.
If, however, we change the variable $\beta$ in the function $f(\alpha, \beta)$ to the current coordinate ${\bf x}$, by transforming the variable using the relations $\beta = {\partial S}/{\partial \alpha}$, as per the Hamilton-Jacobi prescription, we have the transformed function ${\hat f}({\bf x}, t ; \alpha)$, and the Liouville equation (\ref{eq.(1)}) reduces to

\begin{equation}
\frac{\partial {\hat f}}{\partial t} + \frac{\nabla S}{m}\cdot\frac{\partial{\hat f}}{\partial {\bf x}} = 0,
\label{eq.(4)}
\end{equation}
where the momentum term disappears because the momentum variable in the argument here is $\alpha$ and ${\dot \alpha} = 0 $, and where ${\dot{\bf x}}$ in the second term has been written in terms of the momentum as $\nabla S/m$, in consonance with the equation being entirely in terms of the configuation space. Eq.(\ref{eq.(4)}) is now defined only over the configuration space which may be regarded as a Lagrangian submanifold $\mathcal{R}^3$ of the phase space. To appreciate the meaning of eq.(\ref{eq.(4)}), we integrate it over the  momentum variable $\alpha$, which gives

\begin{equation}
\frac{\partial {\cal P}}{\partial t} + \nabla\cdot\left({\cal P} \frac{\nabla S}{m}\right) = 0,
\label{eq.(5)}
\end{equation}
where ${\cal P}$ is the probability density defined by

\begin{align}
{\cal P}({\bf x}) & = \int d\alpha {\hat f}({\bf x}, t, \alpha) , {\rm and} \nonumber \\
 \nabla\cdot({\cal P} \nabla S/m) & = \int {\bf {\dot x}}\cdot\nabla {\hat f}d\alpha =\int (\nabla S/m)\cdot \nabla {\hat f}({\bf x}, t; \alpha) d\alpha ,
\label{eq.(6)}
\end{align}
so that $ {\cal P} \nabla S/m = {\cal P} {\bf v}$ is the probability current.
Eq.(\ref{eq.(4)}) is the equation of continuity as obtained from the Schr\"odinger equation in the WKB approximation, and rightly represents the conservation of probability in the configuration space.

While the underlying dynamics considered so far is still Hamiltonian, a version of the latter, which is more appropriate for the transformed equation, is the Hamilton-Jacobi equation, which is defined entirely over the configuration space $\mathcal{R}^3$. This is

\begin{equation}
\frac{\partial S}{\partial t} + H =0,
\label{eq.(7)}
\end{equation}
with
\begin{equation}
H = \frac{1}{2m}(\nabla S)^2 + V ,
\label{eq.(8)}
\end{equation}
where $S({\bf x}, t )$ is the principal function.

As is well known, through the theory of first order partial differential equations~\cite{courant}, 
the characteristics of the Hamilton-Jacobi equations are essentially the Hamilton equations (\ref{eq.(2)}),

\begin{equation}
{\dot{\bf p}}= - \frac{\partial H}{\partial {\bf x}};\  \  \ {\dot{\bf x}}= \frac{\partial H}{\partial{\bf p}}
= \frac{\partial H}{\partial \nabla S}.
\label{eq.(9)}
\end{equation}
In addition, we have also the relation,

\begin{align}
\frac{dS}{dt}& = \frac{\partial S}{\partial t} + {\bf {\dot x}}\cdot\frac{\partial S}{\partial {\bf x}} \nonumber \\
& = - H + {\bf p}\cdot\frac{\partial H}{\partial {\bf p}} = L ,
\label{eq.(10)}
\end{align} 
as discussed in ~\cite{courant}, which essentially defines the Lagrangian $L$ as the total time derivative of the action $S$, as the principal function, where ${\partial S}/{\partial t} = - H $ and ${\dot{\bf x}}= {\partial H}/{\partial {\bf p}}; {\bf p}= {\partial S}/{\partial {\bf x}}$ have been used in obtaining the last line of eq.(\ref{eq.(8)}) from the previous one. Knowing the solutions of (\ref{eq.(9)}) for ${\bf x}$ and ${\bf p}$ one can then obtain $S$ by a simple integration.

As recognized by Jacobi, the integration of the Hamilton canonical equations is thus equivalent to solving
the partial differential equation which can often be solved more easily by separation of variables. The solutions of the canonical equations can then be obtained by a process of differentiation and elimination.

We now seek to generalize the Liouville equation (\ref{eq.(4)}) defined over the configuration space $\mathcal{R}^3$, by attaching to each point of $\mathcal{R}^3$ the value of the principal function $S$ at that point, defined by the relation (\ref{eq.(10)}), but expressed in units of a small action $\eta$, so that $\Phi = S/\eta $.  The resulting probability function $F({\bf x}, \Phi, t)$ is now defined over the product space $\mathcal{R}^3\times \mathcal{S}^{1}$, where $\Phi$ is now taken to be an angle, to be called the action phase, and $\mathcal{S}^{1}$ is the circle which defines $\Phi$, and which as a manifold represents the $U(1)$ fibre bundle attached to each point of the manifold $\mathcal{R}^3$.

The choice of the action phase to serve as the space $\mathcal{S}^{1}$ is motivated by a remark made by Dirac ~\cite{dirac} that a solution $S$ of the Hamilton-Jacobi equation--the `principal function'--has a greater significance than just providing a solution of the canonical equations through its derivatives. According to him, a solution of the Hamilton-Jacobi equation defines a `family' which ``does not have any importance from the point of view of Newtonian mechanics; but it is the family which corresponds to one state of motion in quantum theory, so
presumably the family has some deep signficance in nature, not yet properly understood".
What Dirac terms a `family', Synge~\cite{synge} designates as a `coherent system of trajectories' and defines them more explicitly. All trajectories emanating from a given space point, but with varying momenta belong to one kind 
of `coherent system'; while all trajectories having the same momenta but varying positions belong to a complementary coherent system or family as per Dirac. Thus the choice of $\Phi = S/\eta$ to represent $\mathcal{S}^{1}$ is a part of the Hamilton-Jacobi formalism.

The incorporation of the principal function in the argument of the probability density function $F({\bf x}, \Phi, t)$ through $\Phi = S({\bf x}, t)/\eta $, would then be in consonance with Dirac's remark and make this function correspond to a state in quantum theory, and which at the same time would represent the one-dimensional manifold ${\cal S}^{1}$, attached to ${\cal R}^{3}$ as per our stipulation.
We have used the unit of action as $\eta$ instead of $\hbar$ as we do not yet want to relate this formalism directly to quantum theory, though ultimately it will be so related in an appropriate manner. Of course, since it is an angle, we would require that the functions defined over it be periodic with respect to it, with a period $2\pi$. The magnitude of $\eta$ to measure $S$ in terms of it, would then be so chosen. It is the confrontation with experiments at an appropriate level that should fix the value of $\eta$.
Furthermore, to make the distribution to correspond to a `coherent system of trajectories', we choose the dependence on the momenta $\alpha$ to be a Dirac measure $\delta(\alpha - \alpha_o)$. 

The equation of evolution of the function $F({\bf x}, \Phi, t)$ is then given by
 
\begin{equation}
\frac{\partial F}{\partial t}+ \frac{\nabla S}{m}\cdot \frac{\partial F}{\partial{\bf x}} + \frac{1}{\eta} \left[\frac{1}{2m}(\nabla S)^2 - V \right]\frac{\partial F}{\partial \Phi}=0 ,
\label{eq.(11)}
\end{equation}
where now the coefficient of the ${\partial F}/{\partial \Phi}$ term is the Lagrangian divided by $\eta$, as it ought to be, but expressed in terms of the momentum, $\nabla S$. 
The dependence of the function $F$ on $ \Phi = S/\eta $ makes it different from the Liouville density function, so that the dynamics represented by eq.(\ref{eq.(11)}) is on the manifold $\mathcal{R}^3\times \mathcal{S}^{1}$, and therefore different
from the classical dynamics, but deterministic nevertheless. This issue will come up for discussion later.

The first point to note about this equation now is that if one
integrates it over $\Phi$ over the interval $(0, 2\pi)$ one would recover eq.(\ref{eq.(4)}) as we identify $\int_{o}^{2\pi}d\Phi F({\bf x}, \Phi, t) = {\hat f}({\bf x}, t)$. The assumption underlying this is that $\eta$ is of such a magnitude that renders the resulting phase such that the function $F$ is periodic in it with a period $2\pi$. However, until we confront experiments, this choice could be taken as arbitrary.

It must now be stated here that Eq.(\ref{eq.(11)}) which has implied in it the $\delta$-function distribution in the initial momenta $\delta({\bf \alpha}-{\bf \alpha_{o}})$, represents the dynamical evolution of 
the `family' rather than of any arbitrary distribution. As such it is 
somewhat different from the standard Liouville equation which describes a flow in the phase space
$({\bf x},{\bf p})$. Eq.(\ref{eq.(11)}), on the other hand, describes an evolution in the space $\mathcal{R}^3 \times\mathcal{ S}^{1} $.

{\bf 3. Hilbert space representation of the equation of evolution for the `family' in the manifold $ \mathcal{R}^{3}\times \mathcal{S}^{1}$  }

The next set of steps are going to be entirely mathematical and are not going to compromise the status of the 
original equation (\ref{eq.(11)}) as representing the deterministic dynamical evolution of the `family',  . The resulting equations obtained at the end of these steps
would then deem to be a description of the dynamics of the `family', though in an entirely different form.

To ensure that $F$ always remains positive definite as it evolves, we write $F=\psi^2$ with $\psi$ a real-valued function. Then $\psi$ will evolve according to the same evolution equation as $F$

\begin{equation}
\frac{\partial \psi}{\partial t} + \frac{\nabla S}{m}\cdot \frac{\partial \psi}{\partial{\bf x}} + \frac{1}{\eta}\left[\frac{1}{2m}(\nabla S)^2 - V \right]\frac{\partial \psi}{\partial \Phi} = 0,
\label{eq.(12)}
\end{equation}
where, as mentioned above, $\eta$ is a unit of action (of sufficiently small magnitude) in terms of which the action 
$S$ is measured to yield the phase $\Phi$. We do not, however, identify it with the Planck quantum, since we do not wish to prejudice the derivation towards a quantum connotation yet. Though meant to ensure the positive definiteness of $F$, the introduction of $\psi =+\sqrt F $ is the first step towards obtaining the desired Hilbert space representation.

Since $F$ was taken to be periodic in $\Phi$ as argued above, so also will be $\psi$. Accordingly, we Fourier decompose $\psi$ in $\Phi$  as follows

\begin{equation}
\psi({\bf x}, \Phi, t) = \sum_{n}{\hat\Psi}({\bf x},n, t)e^{-in\Phi}.
\label{eq.(13)}
\end{equation}

Using it in Eq.(\ref{eq.(12)}) we obtain

\begin{equation}
\frac{\partial{\hat\Psi(n)}}{\partial t} + \frac{\nabla S}{m}\cdot \frac{\partial{\hat \Psi(n)}}{\partial {\bf x}}-
\frac{in}{\eta}\left[\frac{1}{2m}(\nabla S)^2 - V \right]{\hat\Psi(n)}.
\label{eq.(14)}
\end{equation}

This gives a finite time-interval representation of ${\hat\Psi}$ in the form

\begin{equation*}
{\hat\Psi}({\bf x}, n,t) =\exp\left[\frac{in}{\eta}\int_{t'}^{t}dt''\left(\frac{1}{2m}(\nabla S)^2 -V \right)\right]
\end{equation*}

\begin{equation}
{\hat\Psi}({\bf x}-\int_{t'}^{t}dt''(\nabla S)/m,n, t'). 
\label{eq.(15)}
\end{equation}

It is to be noted that all the time integrals in the above equation are along the characteristics and are therefore only functions of the times $( t, t')$, besides that of the initial position ${\bf x_{o}}$. One can thus effect a Fourier decomposition of Eq.(\ref{eq.(15)}) with respect to the spatial variable ${\bf x}$. This yields

\begin{equation}
\begin{split}
{\tilde \Psi}({\bf k}, n, t)= \exp\left[\frac{in}{\eta}\int_{t'}^{t}dt''\left(\frac{1}{2m}(\nabla S)^2 -V\right)\right.\cr
-\left. i{\bf k}\cdot\int_{t'}^{t}dt'' \nabla S/m\right]{\tilde \Psi}({\bf k}, n, t').
\end{split}
\label{eq.(16)}
\end{equation}
Consider next the terms in the exponent which can be rearranged as follows

\begin{equation*}
\frac{in}{\eta}\int_{t'}^{t}dt''\left\{\frac{1}{2m}(\nabla S)^2 -V
-\frac{\eta}{nm}{\bf k}\cdot \nabla S \right\}
\end{equation*}

\begin{equation}
= \frac{in}{\eta}\int_{t'}^{t}dt''\left\{\frac{1}{2m}\left(\nabla S- \frac{\eta}{n}{\bf k}\right)^2 -
\frac{1}{2m}\left(\frac{\eta k}{n}\right)^2 -V \right\}.
\label{eq.(17)}
\end{equation}
Now, introduce a function
\begin{equation}
{\bar \Psi}({\bf k},n, t)= {\tilde\Psi}\exp\left\{-\frac{in}{2m\eta}\int_{o}^{t}dt''\left(\nabla S- \frac{\eta {\bf k}}{n}\right)^2\right\}.
\label{eq.(18)}
\end{equation}

In terms of this function ${\bar\Psi}({\bf k}, n, t)$, Eq.(\ref{eq.(16)}) takes the form
\begin{equation}
{\bar \Psi}({\bf k}, n, t)= \exp\left[-\frac{in}{\eta}\int_{t'}^{t}dt''\left\{\frac{1}{2m}\left(\frac{\eta k}{n}\right)^2 + V \right\}\right]{\bar\Psi}({\bf k}, n, t').
\label{eq.(19)}
\end{equation}

We now take an inverse Fourier transform of Eq.(\ref{eq.(19)}) with respect to ${\bf k}$ which yields
\begin{equation}
\Psi({\bf x}, n, t) = \exp\left[-\frac{in}{\eta}\int_{t'}^{t} dt''\left\{\frac{\eta^2}{2m n^2}\nabla^2 + V \right\}\right]\Psi({\bf x}, n, t'),
\label{eq.(20)}
\end{equation}
while $\Psi({\bf x}, n, t)$ is obtained as the inverse Fourier transform of ${\tilde {\Psi}}({\bf k})$ of Eq.(\ref{eq.(18)}) as below

\begin{equation}
\Psi({\bf x},n , t) = \int \frac{d^{3}k}{(2\pi)^3}{\tilde\Psi}({\bf k}, n) e^{i{\bf k}.{\bf x}} \exp\left[-\frac{in}{2m\eta}\int_{o}^{t} dt' \left(\nabla S- \frac{\eta}{n}{{\bf k}}\right)^2\right].
\label{eq.(21)}
\end{equation}
One immediately gets from Eq.(\ref{eq.(20)}) for infinitesimal time $\tau = t-t'$, the following differential equation for the function
$\Psi({\bf x}, n, t)$

\begin{equation}
\frac{i\eta}{n}\frac{\partial \Psi(n)}{\partial t} = -\left(\frac{\eta}{n}\right)^2 \frac{1}{2m}\nabla^2\Psi(n) + V \Psi(n),\ \, n=1, \ 2, \ 3.....,
\label{eq.(22)}
\end{equation}
which clearly exhibit the Schr\"odinger-form structure of the equations describing the evolution of the functions $\Psi(n)$. We note that, in contrast to the Schr\"odinger equation, we have here an infinite set of equations with $\eta/n$ in role of $\hbar$. 

{\bf 4. Relationship between $\Psi(n)$ and the probability function $F({\bf x}, \Phi, t)$ }

Consider first the relation (\ref{eq.(21)}) between the $\Psi({\bf x}, n, t)$ and the function ${\tilde {\Psi}}({\bf k}, n, t)$ which involves the exponential $\exp[-(in/2m\eta)\int_{o}^{t}(\nabla S- \eta {\bf k}/n)^2]$. The time intgral in the exponent can be written in the form $\sim (\overline{\nabla S}- \eta{\bf k}/n)^{2} t  $. Noting that this exponential has the small parameter $\eta$ in the denominator of its exponent, in the limit of large time interval $t$, this exponential will lead to a vanishing contribution to the integral in eq.(\ref{eq.(21)}), unless $\overline{\nabla S} = \eta{\bf k}/n$. Since $\nabla S $ denotes the particle momentum, this relation identifies $\eta {\bf k}/n $ with the time average of the momentum. 

We shall now examine the connection of the functions $\Psi(n)$, with the initial probability function $ F(\bf x, \Phi, t)$ as charted out through the various transformation relations invoked during the course of the above derivation.  
Now recalling the relation $ F=\psi^2$ through which $\psi$ was introduced and the identification
$\int_{o}^{2\pi}d\Phi F({\bf x}, \Phi, t)= {\hat f}({\bf x}, t)$ we obtain

\begin{equation}
{\hat f}({\bf x}, t)= \int_{o}^{2\pi}F d\Phi  =\int_{o}^{2\pi}\psi^2 d\Phi = \sum_{n}{\hat \Psi}^{*}(n){\hat \Psi}(n),
\label{eq.(23)}
\end{equation}
where we have used the Fourier decomposition (\ref{eq.(13)}) for $\psi$. 

Note, however, that it is the function $ \Psi(n)$, and not ${\hat \Psi}(n)$ which obeys the evolution equation (\ref{eq.(23)}). The former is related to the latter through the transformation relations (\ref{eq.(16)}), (\ref{eq.(18)}) and (\ref{eq.(21)}). Since ${\tilde \Psi}({\bf k}, n, t)$ is the Fourier transform of ${\hat{\Psi}({\bf x}, n, t) }$, one has

\begin{equation}
{\hat f}({\bf x}, t) = \sum_{n}\int \frac{d{\bf k}d{\bf k'}}{(2\pi)^6} e^{i({\bf {k -k'}})\cdot{\bf x})} {\tilde \Psi}({\bf k}, n, t) {\tilde \Psi}^{*}({\bf k'}, n, t).
\label{eq.(24)}
\end{equation} 

Next, using the relation (\ref{eq.(18)}) between ${\tilde \Psi}({\bf k}, n. t)$ and ${\bar \Psi}({\bf k}, n, t)$, one has

\begin{align}
{\hat f}({\bf x}, t)& =\sum\int \frac{ d{\bf k}d{\bf k'}}{(2\pi)^6} e^{i({\bf k}-{\bf k'})\cdot{\bf x}}                                                \nonumber \\
&    \exp\left[-\frac{i n}{2m\eta}\int_{o}^{t}\left\{\left({\bf p}- \frac{\eta{\bf k}}{n}\right)^{2} - \left({\bf p}- \frac{\eta {\bf k'}}{n}\right)^{2}\right\}\right] {\bar \Psi}({\bf k}, n, t) {\bar \Psi}^{*}({\bf k'}, n, t),
\label{eq.(25)}
\end{align}
where we have denoted for simplicity of notation ${\bf p} =\nabla S $.
The exponent in eq.(\ref{eq.(25)}) can be written in the form $[({\bf p}- \eta{\bf k}/n)^2 -({\bf p}-\eta{\bf k'}/n)^2] = (\eta/n)({\bf k'}- {\bf k})[2{\bf p} - \eta({\bf k}+{\bf k'})/n]$. Then changing variables $ {\bf K}=({\bf k}+ {\bf k'})/2, {\bf \kappa} = ({\bf k}- {\bf k'})$, one gets

\begin{equation}
 {\hat f}({\bf x}, t) = \sum_{n}\int\frac{d{\bf K}d{\bf \kappa}}{(2\pi)^6} e^{i{\bf \kappa}\cdot{\bf x}}  {\bar\Psi}({\bf K}+\frac{1}{2}{\bf \kappa}, n, t) {\bar \Psi}^{*}({\bf K }-\frac{1}{2}{\bf \kappa}, n, t) \exp\left[i\frac{{\bf \kappa}}{m}\cdot\int_{o}^{t}dt\left({\bf p}-\frac{\eta{\bf K}}{n}\right)\right]
\label{eq.(26)}
\end{equation}

We note in the exponent in eq.(\ref{eq.(26)}) the time integral $\int_{o}^{t} dt' ({\bf p}- \eta {\bf K}/n) $ may be writren as $ ({\bar{\bf p}}- \eta {\bf K}/mn)t $. In the limit of large time interval $t$, this will lead to a rapidly oscillating exponential factor which will result in a vanishing contribution to the integral in (\ref{eq.(26)}) unless
\begin{equation}
{\bar {\bf p}} = {\bar {\nabla S}}= \eta {\bf K}/n.
\label{eq.(27)}
\end{equation}
This corresponds to the identification of momentum ${\bf p}= \hbar {\bf K}$ in quantum mechanics if we identify $\eta$ with $\hbar$ and choose $n=1$. This is essentially, a repitition of the argument given in the beginning of this section for the relation between $\Psi({\bf x}, n, t)$ and ${\tilde\Psi} ({\bf k}, n, t)$, but now appearing with $\eta{\bf K}/n$ being identified with the time average of the momentum. The origin in the two cases is the same exponential factor.

However, this reduces eq.(\ref{eq.(27)}) to

\begin{align}
{\hat f}({\bf x}, t)& = \sum_{n}\int\frac{d{\bf K}d{\bf \kappa}}{(2\pi)^6} e^{i{\bf \kappa}\cdot{\bf x}}  {\bar\Psi}({\bf K}+\frac{1}{2}{\bf \kappa}, n, t) {\bar \Psi}^{*}({\bf K }-\frac{1}{2}{\bf \kappa}, n, t)   \nonumber \\
& = \sum_{n}\int d{\bf K} W( {\bf K}, {\bf x}, n),
\label{eq.(28)}
\end{align}
where $ W({\bf K}, {\bf x}, n)$ is essentially the Wigner distribution defined by 

\begin{align} 
W({\bf K},{\bf x}, n) &= \int\frac{d{\bf \kappa}}{(2\pi)^3} e^{i{\bf \kappa}\cdot{\bf x}}  {\bar\Psi}({\bf K}+\frac{1}{2}{\bf \kappa}, n, t) {\bar \Psi}^{*}({\bf K }-\frac{1}{2}{\bf \kappa}, n, t)      \nonumber \\
&= \int \frac{d{\bf q}}{(2\pi)^3}e^{-i{\bf K}\cdot{\bf q}} \Psi({\bf x}- \frac{1}{2}{\bf q}, n, t)\Psi^{*}({\bf x}+\frac{1}{2}{\bf q}, n, t),
\label{eq.(29)}
\end{align}
the second line of the above equation representing an alternate form for $W({\bf K}, {\bf x, n})$ obtained by inverse Fourier transform of the function ${\bar\Psi}({\bf K}+ {\bf \kappa}/2){\bar \Psi}^{*}({\bf K}- {\bf \kappa}/2)$.

It is interesting to note that the expression for the probability function ${\hat f}({\bf x}, t)$, which is by definition positive definite and governed by the equation (\ref{eq.(5)}) has been obtained in this formalism as the `marginal' of the Wigner distribution, which has appeared here quite naturally in the process of the above derivation. We also note here that the whole body of manipulations here have been carried out in the $\mathcal{R}^3$-Lagrangian submanifold, with the manifold $\mathcal{S}^{1}$ represented by the angular variable $\Phi$, attached to it. The wave vector ${\bf K}$ appearing in the above expressions represents only the inverse space to the above $\mathcal{R}^3$- submanifold, and cannot represent the momentum variable in the manifold representing the phase space. The `momentum' ${\bf K}$ must therefore be regarded as a `psuedo momentum'. It is therefore not surprising that the Wigner distributions are not positive everywhere, since their information content pertains partly to the $\mathcal{R}^3$-manifold represented by the coordinate ${\bf x}$ and partly to the its inverse manifold; the argument ${\bf K}$ of the Wigner distribution representing a shorter spatial scale of variation of the wave function, and the argument ${\bf x}$, a longer scale 
representing basically the amplitude modulation of the short scale variation. When the two scales are widely separated, the ${\bf K}$ argument can acquire the status of a variable almost independent of ${\bf x}$, and be able to represent the `momentum'. As one knows, the separation of scales is the essence of the WKB approximation. But when the two scales merge, the information contents of the two variables are not independent of each other; the WKB procedure fails.

{\bf 5. The indeterminism of the Schr\"odinger formalism, and the deterministic dynamics represented by the generalized set $\Psi(n)$}

If we choose to normalize ${\hat f}({\bf x})$ such that it is unity normalized all over the space, then we have the probability density

\begin{equation}
{\cal P}({\bf x}, t)= \sum_{n} a(n)\Psi^{*}(n)\Psi(n),
\label{eq.(30)}
\end{equation}
where the $a(n)$ are appropriate coefficients which are consistent with the above normalization but are not determined by the theory. This is because the different equations corresponding to the various $n$ values are uncoupled. As such these amplitudes are independent and may be independently specified. This issue will be discussed in a subsequent section

Equation(\ref{eq.(22)}) along with the relation (\ref{eq.(30)}) constitute the desired Hilbert space representation of the flow equation for the family as described by Eq.(\ref{eq.(11)}). As is clearly seen, Eq.(\ref{eq.(22)}) represent a generalized set of Schr\"odinger form equations with the assumed unit of action $\eta$ appearing in the role of $\hbar$, while Eq.(\ref{eq.(30)}) represents a generalization of the probability prescription ${\grave a}$ la Born to include the various modes $n$.
 
Two points may be noted about the equations obtained above: (i) The starting equation, namely, the equation (\ref{eq.(11)}) of evolution for the family is a first order partial differential
equation describing a flow, while Eq.(\ref{eq.(22)}) are  a set of second order partial differential equations, representing a set of wave amplitudes. 
(ii) Secondly, Eq.(\ref{eq.(11)}) describes a deterministic dynamics since, as emphasized earlier, it represents the dynamics of the `family' though in a somewhat unfamiliar form. Though of the Schr\"odinger-form they must collectively describe a deterministic dynamics, since no information is lost in the process of a mathematical
transformation from Eq.(\ref{eq.(11)}) to Eq.(\ref{eq.(22)}). These wave amplitudes essentially represent the coefficients of expansion in terms of the state vectors belonging to the circle $ S^{1}$. It can now be seen that the complete set of state vectors would together specify the deterministic dynamics contained in the set of equations (\ref{eq.(22)}) which mirror the deterministic dynamics as defined by the flow equation (\ref{eq.(11)}) in the higher dimensional product space ${\cal R}^{3}\times S^{1}$. The selection of only one of these modes--$n=1$--to describe the dynamics would compromise the completeness of the set of state vectors, and would render the dynamical description incomplete and thereby probabilistic.

We shall discuss in the following the relation between the probabilistic description as afforded by the Schr\"odinger equation and the deterministic description 
as represented collectively by the set of equations (\ref{eq.(22)}).

{\bf 6. The generalized set of Schr\"odinger form equations, quantum mechanics and the classical limit}
 
Given the fact that our starting equation (\ref{eq.(11)}) is a deterministic flow equation in the space $\mathcal{R}^{3}\times \mathcal{S}^{1}$, the resulting set of transformed equations--the Schr\"odinger-form equations, describing a Hilbert space representation--
must together describe a deterministic dynamics, because, as alluded to above, no information is lost in the process of this transformation. Of course, this determinism is not to be confused with the determinism of classical mechanics. Here we have a different dynamics.

However, we do recover the Schr\"odinger equation as the $n=1$ mode. To that extent
this does seem to vindicate Dirac's conjecture that the family of solutions which is represented by a solution of the Hamilton-Jacobi equation has a close relationship with quantum mechanics, if we do identify $\eta$ with $\hbar$. But as one knows, the Schr\"odinger description is known to be intrinsically probabilistic and thus `incomplete'
from the standpoint of a deterministic description. We are thus led to the question whether the exclusion of the rest of the modes $n=2, \ 3, \ 4, ....\infty $ from the description compromises the determinism which all the modes together would appear to guarantee. A deterministic dynamics implies a $\delta$-function for the path (which is a trajectory determined by the equations for the characteristics), which in turn requires equal weights for the various Fourier components.

If one takes the point of view that  Nature makes a selection of the $n=1$ as the only one it wants to manifest itself with,--Nature's fiat-- then clearly the resulting description will necessarily be `incomplete' from this standpoint and therefore intrinsically probabilistic. The well known indeterminism of quantum mechanics may then be attributed to this incompleteness.

However, if one takes the point of view that other modes $ n= 2, \ 3, \ 4,...$ may also manifest, though not with equal weights; then in that case these modes could be regarded as representing new modes of non-classical behaviour. Even in that case the `in principle' determinism implied by the generalized flow equation (\ref{eq.(11)}) stands compromised.
On the other hand, it may also be noted that the modes with large $n =N$ leads to $\eta/ N \rightarrow 0 $. Using this limit, the various large $N$ modes lead in the WKB- approximation to the classical limit--the Hamilton-Jacobi equation. On could thus say that the large $ n$ modes, $N>> 1$ contribute to the classical behaviour, while $n=1$ describes the quantum behaviour. Thus the two kinds of behaviours are recovered in the two opposite regimes. It is in this sense that the present formalism covers both classical and quantum mechanics. However, it is pertinent to remark that it is not necessary now to take the formal limit $\hbar\rightarrow 0$ to obtain the classical description. This is accomplished here by the $N\rightarrow large $ limit. In fact, for a lagre $N$, the corresponding equation can be written in the form

\begin{equation}
i\eta\frac{\partial \Psi(n)}{\partial {\bar t}} = - \frac{\eta^2}{2m}{\overline \nabla}^2\Psi(n) + V \Psi(n),
\tag{30A}
\end{equation}
where we have written ${\bar t} = N t$ and ${\overline \nabla}^2 = [\nabla/N]^2\sim 1/L^2 $, so that both the time variable ${\bar t}$ and the spatial variable ${\bar {\bf x}}= N{\bf x} $ represent variables on the longer scales in time and space and thus characterize classical regime. If we balance the term on the left of eq.(30 A) with the first term on the right, then the latter has the coefficient of magnitude $\sim \eta T/(mL^2) \sim \lambda_{dB}/ L $; (if we identify $\eta $ with $\hbar$, and thus write $\eta(T/Lm) =\lambda_{dB}$, $\lambda_{dB}$ being the de Broglie wave length.) If, as stipulated, $L$ represents a macroscopic length, then the coefficient $\lambda_{dB}/ L $ is indeed $<< 1$ and thus 
characterizes the classical limit. Thus looked at this way, the large $N$ characterizes the classical behaviour.

We shall now consider the experimental implications of the existence of other modes of behaviour corresponding to $ n=2, \ 3, \ 4, .... $. Could they represent additional modes of  non-classical behaviour? This may appear to be a rather heretical proposition at first sight, in the face of such enormous successes of quantum mechanics and with no departures observed so far.
Be that as it may, it is nevertheless interesting to look for such possibilities, for at times new effects may be lurking and may not manifest for lack of a deliberate search.
 
\vskip 5mm
{\bf Two possibilities present themselves:} 
\vskip 5mm
One refers to the implications of the new modes with reference to the quantum tunneling of  potential barriers, and the other pertains to matter wave interference in a double slit experiment or Bragg reflection of electrons from periodic lattice of crystals. In both the cases the phenomena involve direct interaction of electrons with electrostatic potential rather than via the photon field as in the case of spectral lines from an atom. There is, therefore, a greater likelihood of manifestation of the new modes, if they exist, in the two cases mentioned above.

{\it 6.1. Quantum tunneling of potential barriers}

One of the most direct experimental demostration of the quantum tunneling is presented by the beautiful experiments by Binnig et al.~\cite{binnig} on the tunneling across a vacuum gap which serves as a potential barrier for electrons going from one metal to another. The theory for this effect is an elementary exercise in quantum mechanics. But the experimental demonstration required some highly skilful experimentation. The experiments showed that the tunnel current across the vacuum gap decreased exponentially with the width $s$ of the gap, as revealed through the linearity of the ${\rm ln}\tau \ vs \ s $ plot. This is, of course, as expected from quantum mechanics. However, some of the experimental plots presented in the paper~\cite{binnig} reveal a behaviour which is difficult to understand in standard terms. A single exponential dependence of the electron current on the gap width would lead to a straight line on a semi-log plot. That is what is indeed observed for the major portion of the tunnel current over about three orders of magnitude as reported in the paper ~\cite{binnig}. But two of the curves in the plot, which correspond to cleaner surfaces of the metallic samples, exhibit a non-linear dependence over a distance of about an Angstrom in the part corresponding to the tip being closest to the surface, indicating the possible existence two exponential terms instead of just one. A break is clearly discernible in the linearity in the higher current region of the plot at about $\sim 5. 10^{-7} amp$.
An estimate carried out by the author shows that the slope of this portion of the curves is approximately twice that of the major part of the plot. In fact, the two curves designated as $D$ and $E$ as shown in the plot in the reference ~\cite{binnig} are seen to be represented by the fits given below

\begin{align}
Plot D \Rightarrow I_{D} &=  0.116 \times 10^{-2} e^{-\kappa_{1} (\Delta x)}\cr 
&+  2.26 e^{-\kappa_{2} (\Delta x)}, \  \  \kappa_{1} = 1.745, \kappa_{2}=3.5 \nonumber \\
Plot E \Rightarrow I_{E} & =  0.22 \times 10^{-4} e^{-\kappa_{1} (\Delta x)} \nonumber \\ 
& +  0.735 \times 10^{-3} e^{-\kappa_{2}(\Delta x)}\   \   \  \nonumber \\
\kappa_{1}&=1.72, \kappa_{2}=3.4
\label{eq.(31)}
\end{align}

We ought to qualify that the minimum value of $\Delta x$ which corresponds to the maximum tunnel recorded
 ($ 10^{-6}$ amp) is $\Delta x = 4.4 \AA $ for the plot D, while for the plot E it is $\Delta x = 2.17 \AA $, the $\Delta x$ being the distance as indicated on the location of the curves in the plot and not the width of the actual gap. The widths $s_{d, e}$ of the actual gaps in the two case are thus $s_{d}= (\Delta x - 4.4)\AA,\  \ s_{e} =(\Delta x - 2.17)\AA $.

It is interesting to note that in both cases ( Plot D and E) the ratio $\kappa_{2}/\kappa_{1}\simeq 2 $. In fact, for the Plot D, $\kappa_{2}/\kappa_{1}= 2.01$ while for the Plot E, $\kappa_{2}/\kappa_{1} = 2.0 $ . The second exponential in the tunnel current would thus appear to correspond to the mode $n=2$ of our set of equations
(\ref{eq.(22)})with $\eta$ having been identified with $\hbar$. Equally interestingly, the share of both the modes $ n=1,n=2 $ in both the cases D and E is seen to be almost equal at the highest recorded current of $10^{-6}$ amp. For the case D it is $I^{D}_{1} = 0.537 \times 10^{-6}{\rm amp}, I^{D}_{2}= 0.463 \times 10^{-6}{\rm amp} $. For the case E it is {$ I^{E}_{1}= 0.526 \times 10^{-6}{\rm amp}, I^{E}_{2}= 0.474 \times 10^{-6}{\rm amp}$, almost the same in both the cases. This is a rather significant fact, that the current corresponding to the mode $n=2$ competes with the one for $n=1$, at the highest recorded current and cannot be ignored. Their is an unmistakable presence of nonlinearlty in both the curves $ D$ and $E$, and which is seen to correspond to the presence of a second exponential with double the exponent. The authors of ~\cite{binnig} have not commented on this nonlinearity in the observed tunnel current. 

Unless an alternative mechanism is available to explain this component resulting in the nonlinear departure, it offers a strong possibility as a manifestation of $n=2$ mode. On the other hand, a case can be made for carrying out these experiments more specifically and carefully to check whether such a behaviour can be reproduced in the experiments, which can then be studied more extensively, or whether this effect can be traced to any artefact of the experiment.

We consider next the implications of the new modes for the matter interference effects as predicted by these modes.

{\it 6.2 The double slit experiment}

Consider the double slit experiment with the two slits in a screen in the $y-z$ plane at distances $y= \pm d $,
with the observation screen placed at a distance $x=X$ from the plane of the slits. Let the slit at $y= +d$ be designated as slit $`1'$, while that at $y= -d $ be called slit $`2'$. Let the particles emanating from $`1'$ have a small spread with a distribution $\exp[-\beta(y-d)^2]/|{\bf r}- d {\hat{\bf e}}_{y})|^{1/2}$, and those emanating from the slit $`2'$ with the distribution $\exp[-\beta(y+ d)^2]/|{\bf r}+ d {\hat{\bf e}}_{y}|^{1/2}$, where ${\bf r}\equiv (x, y)$ is the radius vector from the origin to any field point. The denominators in these distributions describe the `spreading out' of the waves from the line of origin in the two-dimensional system being considered. These spreads may be regarded as arising from a wave packet formation when the waves emanating from the slits have a small wave number spread, $\Delta{\bf k}$ around a mean wave number ${\bar{\bf k}}$. The wave amplitudes of these wave packets are then superposed as they reach a point ${\bf r}$ on the
observation screen and are given by the sum

\begin{eqnarray}
\psi({\bf r})& = & \frac{A_{o}}{|{\bf r}-d{\hat{\bf e}_{y}|^{1/2}}} e^{-\beta(y- d )^{2}}  e^{i{\bf k}\cdot({\bf r}-d{\hat{\bf e}}_{y})}\cr
& +& \frac{A_{o}}{|{\bf r}+d{\bf {\hat e}}_{y}|^{1/2}} e^{-\beta(y+ d )^{2}}  e^{i{\bf k}\cdot({\bf r}+d{\hat{\bf e}}_{y}) }.
\label{eq.(32)}
\end{eqnarray}

The probability density is then given by (taking mod-squarred)
\begin{eqnarray}
|\psi({\bf r})|^2 &= &A_{o}^2\left[\frac{e^{-2\beta\left(y-d\right)^2}}{|{\bf r}-d {\hat{\bf e}_{y}|}}+ \frac{e^{-2\beta\left(y+d\right)^2}}{|{\bf r} + d{\bf {\hat e}_{y}}|} \right]\nonumber \\
&+ & 2 A_{o}^2\frac{ e^{-2\beta(y^2+ d^2)}}{(X^2 + y^2 -d^2)^{1/2}}\cos 2{\bf k}\cdot {\bf d}.
\label{eq.(33)}
\end{eqnarray}
The first two terms of the expression (\ref{eq.(33)}) represent distributions on the screen centered around the points $ y= \pm d $, the $\pm$ signs defining directions in lines with respectively the slits `1' and `2'. These distributions correspond to the expectation of the classical behaviour. The third term on the other hand
corresponds to the interference between the trajectories emanating from the two slits, and expresses quantization
of the direction through the cosine factor $ \cos(2{\bf k}\cdot{\bf d})$. The maxima of the cosine terms come from

\begin{eqnarray}
2{\bf k}\cdot{\bf d}& = & 2\ell\pi, \  \ \ell= 1,\ 2,\ 3,..... \cr
2 k d  \sin \varphi & = & 2\ell \pi ,
\label{eq.(34)}
\end{eqnarray}
where $\varphi$ is the angle which the radius vector to the field point $(X, y)$ makes with the x-axis. This gives the quantization of direction, which is the signature of the interference fringes.

We next consider the effect of the inclusion of the other modes $n=2,\ 3, \ 4, ....$.  Including all the modes $n$, we have the following expression for the probability density $|\Psi(n)|^2$ summed over all the modes $n$ which can be written down from the expression (\ref{eq.(33)})
\begin{eqnarray}
\sum_{n}&|\Psi(n)|^2 &= \nonumber  \\
&\sum_{n}&\frac{{A_{o}^{(n)}}^2}{(X^2 + y^2)^{1/2}}\left[e^{-2\beta (y-d)^2}+ e^{-2\beta (y+d)^2}\right]\nonumber \\
&+ & \sum_{n} \frac{{A^{(n)}_{o}}^2}{(X^2 + y^2)^{1/2}} e^{-2\beta( y^2+ d^2)}2\cos 2n{\bf k}\cdot {\bf d}.
\label{eq.(35)}
\end{eqnarray}

where we have neglected $ d^2 << X^2 $ in the denominators of the above expressions.

To study the dependence of the probability density on the relative magnitudes of the various terms in the above summation over an infinite number of $n$ values we take
$A^{(n)}_{o}$ to vary with $n$ with a negative exponential as $A_{o}\exp[-\alpha( n-1)/2]$, with $\alpha$ being an arbitrary appropriate positive number. This form ensures that the mode $n=1$ is unaffected by the value of $\alpha$. Carrying out the sum we have

\begin{align}
\sum_{n}|\Psi(n)|^2 &= \frac{{ A_{o}}^2}{(X^2 +y^2)^{1/2}}\left[e^{-2\beta (y-    
                                                 d)^2}+ e^{-2\beta (y+d)^2}\right]    \nonumber \\
                               & +  \frac{{ A_{o}}^2}{(X^2 + y^2)^{1/2}} e^{-2\beta( y^2 + d^2)}\frac{2(\cos\theta  - e^{-\alpha})}{2e^{-\alpha}(1-\cos \theta)+(e^{-\alpha}-1)^2},
\label{eq.(36)}
\end{align}
where $\theta \equiv 2{\bf k}\cdot{\bf d}$, and where the sum over $n$ has been evaluated for a finite value of
$\alpha$, with the large value $N$ for the upper limit for the summation index $n$.  We notice from the above expression that in the limit $\alpha\rightarrow \infty$ the factor giving the $\theta$ dependence of the interference term reduces to $\cos\theta$. This corresponds to the $\theta$-dependence of the interference term in the quantum mechanical expression (\ref{eq.(33)}). This is consistent with our stipulation that in the limit $\alpha\rightarrow \infty$ only the term $n=1$ contributes which corresponds to the expectations from quantum mechanics.

{\it 6.3 Approach to the classical behaviour}

However, a fascinating case arises when $\alpha = 0 $,  and  $N\rightarrow \infty $. This corresponds to all the various $n$-modes contributing with equal weights. In that case the sum over the interference term in eq.(\ref{eq.(35)}) must be evaluated separately. This is found to be

\begin{align}
lim_{N\rightarrow \infty}\sum_{n=1}^{N} \cos n\theta &= -1 + \frac{\sin(N+\frac{1}{2})\theta}{\sin \frac{1}{2}\theta} \nonumber\\
& = -1 + 2\pi \delta (\theta)
\label{eq.(37)}
\end{align} 

In that case the last term in eq.(\ref{eq.(35)}), that represented the interference term, reduces to
$ {A_{o}}^{2} \exp{[-2\beta(y^2+ d^2)]}(-1+ 2\pi \delta(\theta))$, which exhibits no undulatory $\theta$ dependence characteristic of the interference effects. In fact, since $\theta \equiv 2{\bf k}\cdot{\bf d}$, the only direction that is allowed by virtue of the delta function $\delta(\theta)$ is given by $\theta\equiv 2{\bf k}\cdot{\bf d}=0$. This implies the wave vector ${\bf k}$ being normal to the plane of the slits; which means that only the direction of the 
initial wave vector in the line with the slits is permitted, the particles along which contribute to the classically expected humps on the observation screen. If one now integrates the interference term formally over $\theta$ so as to extract the contribution of $\delta(\theta)$ then one has $\int_{-\pi}^{\pi} d\theta [-1 + 2\pi \delta(\theta)] =0 $. Thus the interference term vanishes, leaving only the two humps being described by the probability distribution

\begin{align}
\sum_{n}|\Psi(n)|^2 &= \frac{{ A_{o}}^2}{(X^2 + y^2)^{1/2}}\left[e^{-2\beta (y-d)^2}+ e^{-2\beta (y+d)^2}\right]. 
\label{eq.(38)}
\end{align}

This is essentially a distribution expected in accordance with the classical behaviour--just the two humps in the line of the slits.   

It may now be remarked that the summation over the whole range of $n$ amounts to integration of the function $ F({\bf x}, \Phi, t)$ over the variable $\Phi$, which yields the distribution ${\hat f}({\bf x}, t)$ as expressed by eq.(\ref{eq.(23)}), and which obeys the classical Liouville equation (\ref{eq.(4)}) in the Lagrangian submanifold $\mathcal{R}^3$.
Thus, the classical behaviour is recovered when subject to identifying the momentum vector $ {\bf p}$ with $ \eta {\bf K}/n $, the expression is integrated over ${\bf K}$, as expressed by eq.(\ref{eq.(28)}).

This formalism thus provides a different way to look at classical--quantum relationship. In the standard view, the classical limit of the interference pattern was supposed to be obtained by smearing over the closely spaced interference fringes which are supposed to result with massive objects in the double slit experiment, because of the associated short de Broglie wave length, as for instance, described by Feynman~\cite{feynman}. In the present picture, the interference pattern disappears entirely because of the summation over the modes $n$. It may be remarked that the two pictures mentioned above correspond to different forms for the limiting distribution on the screen. In the picture ${\grave a}$ la Feynman, there is a hump centered around of the mid- point of the screen, while according to the picture represented by (\ref{eq.(37)}), there are two humps in line of the slits. The latter picture would appear to be more in tune with the classical expectation.
 
{\it 6.4  Possible new modes of quantum behaviour in the double-slit experiment }

Assuming that there is no Nature's fiat to rule out strictly all the modes other than $n=1$,  one can then consider the existence of other modes as well, and their experimental consequences. As we have seen that if $\alpha$ appearing in eq.(\ref{eq.(36)}) is sufficiently large, then only the $n=1$ mode survives in the limit. This is equivalent to Nature's fiat.
However, if $\alpha$ is not large enough, then the other modes corresponding to $n=2, 3,..$ could also manifest in the inteference fringes in the form of harmonics of the fundamental $n=1$. To check whether such harmonics are present even in small amounts, we have looked at the double-slit inteference fringes obtained in the experiments with single electrons reported by Hitachi Inc. on the web, which exhibit the remarkable building up of the interference fringes as the electrons from an electrostatic biprism are recorded while they arrive on the screen one by one. The reason for chosing these experiments for our analysis is that the prcoess involved in the building up of the interference fringes would be direct and quite clean, uncontaminated
by the possibilities of other effects which could be present in systems, such as crystals, where the presence of harmonics could be attributed to effects such as the multiple scattering from other crystal planes.

We have scanned these figures (two of them) to convert them into plots depicting the intensity variation with respect to the transverse position across the screen. These
two plots are depicted in Fig. 1 in the two adjacent frames A and B. These plots are then subject to a Fourier decomposition to determine if there exist harmonics in them corresponding to n=2, 3, etc. as discussed above, expected as per the formalism presented here. The Fourier plots so obtained are presented below the respective curves in the two frames. The axes bear arbitray units because we only need to identify the presence of harmonics and their relative intensities.

\begin{figure}[!ft]
\includegraphics[width =8 cm]{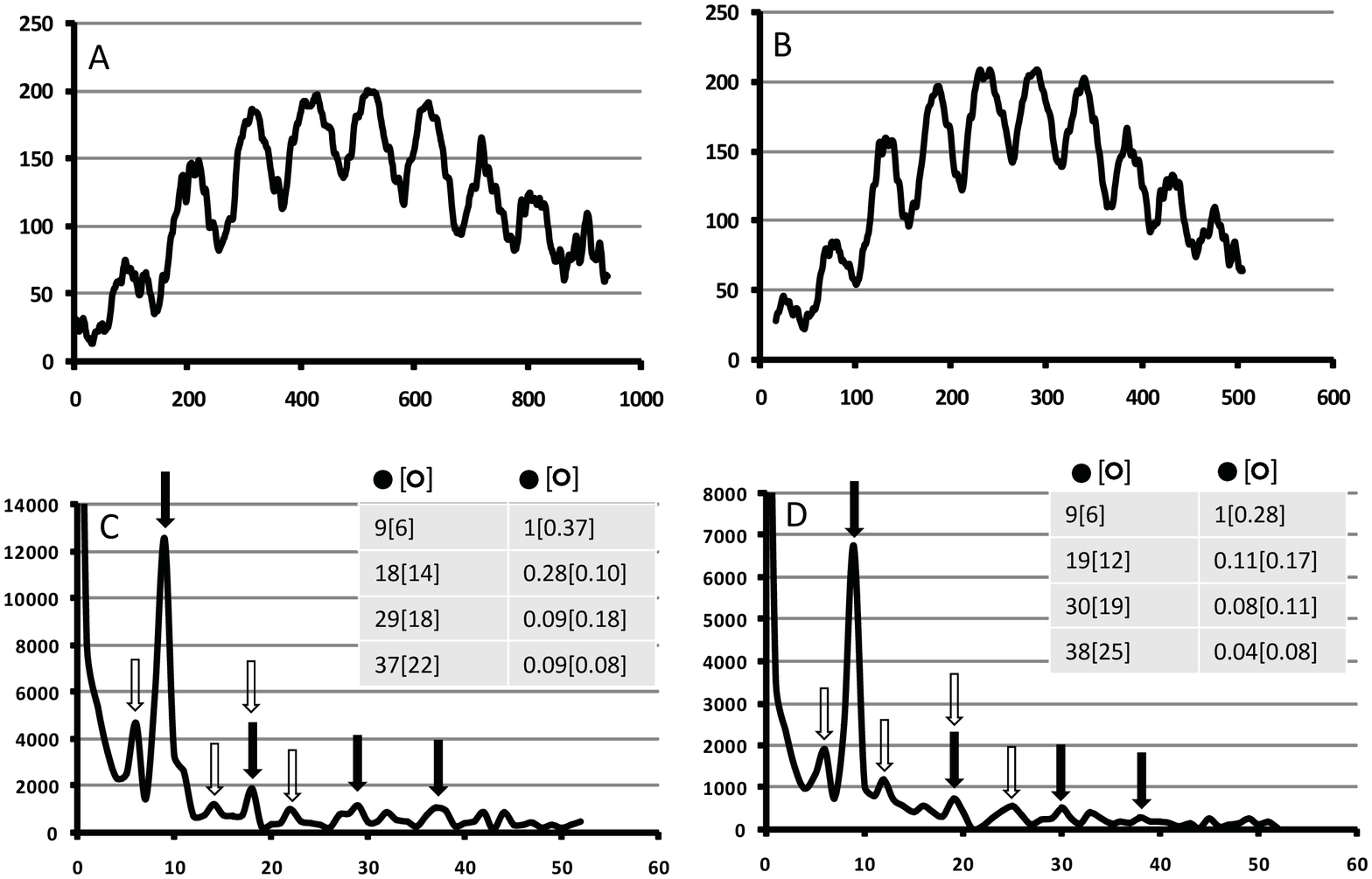}
\centering
\caption{\small{\bf Interference fringes in the double slit experiment} The top frames (A and B) in the figure depict the actually observed interference fringes scanned for the intensity across the fringes. The bottom frames(C and D) give the Fourier plots of the observed intensity profiles. The plots clearly show the presence of dominant  peak corresponding to the $n=1$ mode at the frequency $f= 9$ in arbitrary units. The other subdominant peaks at the harmonics $n=2, 3, 4, ...$ are clealy present. These are shown in the figures as `solid' arrows with the frequencies given in the bottom frames. Also seen to be present are a set of harmonics peaks with a smaller fundamental frequency at $f=6$ and with smaller intensities.These are indicated with `hollow' arrows. The numbers against the frequencies give the relative intensities of the peaks normalized to the lagest peak corresponding to $f=9$}
\label{Fig. 1}
\end{figure}

A dominant peak is clearly visible in both the Fourier plots at the same position $f=9$ on the frequency axis. But it is interesting to notice the presence of (sub-dominant) harmonic peaks as well which are indicated by the vertical arrows. We can easily identify the presence of up to $ n=4$ peaks in the plots.

Notice, however, also the presence of some other sets of peaks-- much smaller than the aforementioned ones--which also form a harmonic sequence, with the frequency of its $ n=1$ peak being a little smaller than that of the dominant sequence. These are indicated by `hollow' arrows, as against the `solid' ones for the dominant sequence. We do not understand the origin of these sub-dominant set of harmonic peaks. There may possibly be present in the electron beam a small, somewhat lower energy component which is responsible for this other sequence. The interesting thing, however, is that they too form a harmonic sequence. These sequences for both the dominant and sub-dominant sequences are indicated
in the inset in the two frames A and B, which give in the left column `frequencies' corresponding to the harmonics for the two 
sequences- with the numbers in the square brackets giving the frequencies for the sub-dominant sequence.[Frame C-9:18:29:37 :: 1:2:(3.2):(4.1), Frame D-9:19:30:37::1:(2.1):(3.2):(4.2)]. The figures in the square brackets refer to frquencies of the sub-dominant sequence of peaks, which too form a harmonic sequence[ Frame C-
6:14:18:22 :: 1:(2.3):(3):(3.7), Frame D-6:12:19:25 :: 1:2:(3.1):(4.1)]. The right column on the other hand give the `amplitudes' of the different frequencies relative to the one for the dominant peak in both the cases.

The presence of up to four harmonics is clearly discernible in both the frames C and D. Since the experimental results analysed pertain to a direct, particle by particle build-up of the interference fringes in the double prism experiment, it ought to be regarded as free from any artefact. Consequently, the presence of harmonics, as expected from the proposed formalism, ought to be regarded as an indication of the validation of the formalism, unless some other explanation can be found for the presence of such harmonics. On the other hand, it would be desirable to carry out such experiments more carefully so as to specifically look for these harmonic peaks

{\bf 7. Concluding comments}

The Schr\"odinger formalism of wave mechanics has been the cornerstone for the development of the entire gamut of physical phenomena on the micro-scale in the non-relativistic limit. As one knows well, this equation was obtained by Schr\"odinger using the relationship between the wave optics and the ray optics which is described by the eikonal equation representing the geometrical optics limit to the wave equation. Recognizing that the Hamilton-Jacobi equation of classical mechanics would be the geometrical optics counterpart of a wave equation that a de Broglie wave would follow, he was led to his wave equation through a certain procedure. The actual procedure, though interesting, is essentially of historical interest today.

It may be recalled that the Born interpretation of the wave function as a probability amplitude was essentially postulated later. 

The formalism proposed here, which includes the Schr\"odinger equation as a special case, obtains a generalized set of Schr\"odinger equations, as a Hilbert space representation of the flow equation representing the evolution of a `family'. One has thereby obtained a derivation of the Schr\"odinger formalism in a generalized form where the probability interpretation follows from the derivation itself, as against the prescription of Born in the conventional theory.

The mathematical procedure used in the construction of the above mentioned Hilbert space representation is similar to the one used by the author in connection with the problem of charged particle dynamics in a magnetic field, where a similar set of equations were obtained starting from the classical Liouville equation for the system ~\cite{rkv(85)}. These equations had led to some very fascinating predictions about the behaviour of the system, which are found to be characteristcally matter wave-like, and which have since been comprehensively verified in a series of experiments ~\cite{rkv-p-b, rkv-b, rkv-b-a}. The procedure used here thus stands validated.

Even if the formalism presented here did not yield anything other than the known Schr\"odinger formalism, then it would still be interesting for the manner in which it is obtained, and because of the general framework in which the Schr\"odinger formalism is seen to be embedded. But we do find something more:
We find other modes of the propagation of probability, namely, those corresponding to $n=2,\ 3, \ 4, ....$. As we have demonstrated here, there seems to be some preliminary evidence of the existence of some of these modes in the two situations: (i) electron tunneling across vacuum gaps, where we have identified a tunnel current corresponding to the $n=2$ mode, (ii) double-slit interference fringes where we have identified the existence of harmonic peaks upto $n=4 $, corresponding to these modes. 

Both the situations were chosen as to correspond to a direct interaction of electrons with the different forms of potential, rather than those involving any intermediate field like the photon field as in the case of spectral lines. These results are admittedly preliminary, and more specific experiments must be carried out to check whether these results stand confirmed, and whether there is an alternate explanation for them, or whether they represent any artifact in the experiments. Nevertheless, they do provide the motivation that efforts ought to be directed to carry out such experiments more exhaustively.

There are two important issues which motivated this investigation. One was the question of the probabilistic nature of the quantum mechanics and the so-called `incompleteness' of its description ${\grave a}$ la Einstein. The whole `hidden variable'
programme of the last few decades was designed to `rectify' this by supplementing the systems on the micro-scale with a local hidden variable space, which was meant to remedy the `incompleteness'. The hidden variables were attached to the wave functions so as to render them dispersion free. This approach, however, failed to attain the desired objective, since certain relations --the Bell inequalities--based on the local hidden variable theories, were found to be in inconflict with experiments and whose results agreed with quantum mechanics.

In the present picture the determinism is effectuated not by an ad hoc supplementation of the wave function by a hidden variable space, but in a somewhat more fundamental way by extending the configuration space over which the Schr\"odinger wave functions are defined  by attaching to it a $S^{1}$ manifold. The $S{1}$-manifold is also not any arbitrary one, but the one pertaining to the system itself, namely the action phase defined by the principal function for the system. A deterministic dynamics is then defined over this product space. The determinism here exists in this extended space, and is represented by all the infinity of $n$ modes in the space $\mathcal{R}^3\times {\cal C}$, when this dynamics is transformed in terms of the elements of the vector space represented by the circle $ S^{1}$. The incompleteness of the Schr\"odinger description is then related here to the absence of the other modes $n= 2,\  3, \,....\infty $, either through Nature's fiat, or through any other not-too-clear a mechanism which could attenuate those modes.

There seems to exist preliminary evidence that Nature's strict fiat may not be operating. The evidence, however preliminary, is seen to exist in two cases discussed in detail in Sec.
6.1 relating to quantum tunneling across vacuum gaps, where a tunnel current correesponding to $n=2 $ mode has been identified, and in Sec. 6.4, where the presence of harmonics in the interference fringes corresponding to the modes upto $n=4 $ have 
been identified.

The other motivation for this investigation was to obtain a formalism that represents a covering wave amplitude representation, which yields both quantum and classical dynamics in different limits. As discussed above, this formalism yields quantum mechanics strictly for $ n=1$ and some deviations from it for small mode numbers $n= 2, \ 3, .....$; and yields classical dynamics in the large mode number limit $n \rightarrow \infty $, in the WKB spirit. What is interesting to note is that it is not required for the WKB limit here to have the Planck quantum $\hbar$ take the formal limit $\hbar \rightarrow 0$. Rather, the limit of large $n$ takes care of the required limit to yield classical dynamics in the WKB spirit. 

However, an interesting feature of this formalism that we have found, is the recovery of a classically expected pattern in the double slit experiment when all the infinite $n$ modes are included in the expression of the probability density. The interference term still exists, but it is no more undulatory as in the observed double-slit experiment. In fact, this interference term is non-zero only for wave vectors normal to the slit plane. This means that no scattered waves from the slits are permitted. This leaves only the two humps in the line of the slits, charactersitic of the classical behaviour. On the other hand, the usual double-slit undulatory inteference term is obtained if only the $n=1$ term is operative. It would therefore seem as if the indeterminism of quantum mechanics could be attributed to the selection of only one of the infinity of modes--the $n=1$ mode--to the exclusion of others. From the point of view of this picture, it means that the interference effects are intimately related to this indeterminism--a conclusion which could bring into question the assignment of meaningful trajectories in the double-slit interference experiment, as is done in some theories. The inclusion of these other modes with equal amplitudes yields the deterministic classical behaviour. 

The indeterminism would still prevail even if some of the lower order modes $ n=2, \ 3, \ 4, ...$ are found to exist if the observations of some these modes as pointed out above are found to be confirmed; because the complete determinism would require the specification of all the modes with equal amplitudes.

We have here considered only a one-particle system; so the questions of distant quantum correlations of the EPR type and quantum entanglement do not confront us here. However, the formalism developed here can be applied to a many particle system which would require the product space $\mathcal{R}^{3N}\times S^{1}$ to build it on, so that the manifold $S^{1}$ is attached to the $ 3N$-dimensional configuration space manifold. It is this fact which will be found to be related to the quantum correlations. The uncorrelated many particle system in this context will be, on the other hand, represented by $\prod_{i=1}^{N}\mathcal{R}^{3}_{i}\times S^{1}_{i}$ The corresponding formalism and its implications will be investigated later. 

The formalism presented here is a first step towards a different point of view. This will need to be developed further to explore many other important issues relating to classical- quantum relationship as well as quantum mechanics itself.
 
{\bf Acknowledgment} This work is supported by the Platinum Jubilee Fellowship of the National Academy of sciences, India, which is gratefully acknowledged.

\end{document}